\documentclass[twoside]{article}
\usepackage{amsthm,amsmath}
\usepackage{amssymb}
\usepackage{bm}
\usepackage{bbm}
\usepackage{url}
\usepackage{url}
\usepackage{graphicx}
\usepackage{color}
\usepackage{nicefrac}
\usepackage{tikz}
\usepackage{graphicx}
\usepackage{caption}
\usepackage{float}
\usepackage[makeroom]{cancel}
\usepackage[all]{xy}
\usepackage{subcaption}

\usepackage[accepted]{aistats2015}
\usepackage[aboveskip=0pt]{caption}
 
\definecolor{purple}{RGB}{191,85,236}

\begin{document}

\twocolumn[

\aistatstitle{Criticality \& Deep Learning II: Momentum Renormalisation Group}

\aistatsauthor{ Dan Oprisa
\\ \texttt{dan.oprisa@critical.ai} \And Peter Toth\\ \texttt{peter.toth@critical.ai}}

\aistatsaddress{ \\CriticalAI \\ http://www.critical.ai } 
]

\begin{abstract}

Guided by critical systems found in nature we develop a novel mechanism consisting of inhomogeneous polynomial regularisation via which we can induce scale invariance in deep learning systems. Technically, we map our deep learning (DL) setup to a genuine field theory, on which we act with the Renormalisation Group (RG) in momentum space and produce the flow equations of the couplings; those are translated to constraints and consequently interpreted as "critical regularisation" conditions in the optimiser; the resulting equations hence prove to be sufficient conditions for - and serve as an elegant and simple mechanism to induce scale invariance in any deep learning setup.

\end{abstract}

\section{Introduction}

The ubiquity of self similarity stemming from universal scale invariant behavior displayed by virtually all systems in various disciplines serves as motivation of the current research; 
starting from biological systems \cite{origin_order,biology_poised}, including the brain \cite{brain_crit}, physical systems \cite{fisher_renorm} and even on large cosmological scales \cite{cosm_scale} self similarity is encountered. There are various underlying mechanisms producing the emergent scale invariance, some of which rely on tunable parameters \cite{scale_renorm_univers, diff_eq} and some of which are self-organised \cite{per_bak}. Hence the conjecture is near, that self similarity, scale invariance, power law distribution, criticality are all just facets, emergent patterns of underlying symmetries at heart of complex systems. In the following we will also treat those interchangeably as they are just aspects of a deeper underlying structure.

The brain on the other side, is arguably one of the most complex systems known to us, displaying architectural and functional level power law patterns. Given the universality of power law behavior and the biological findings about the brain, it seems almost necessary to consider those emergent laws as a necessity for intelligence and hence incorporate them (or the underlying generating mechanism) into the respective DL systems. 

In order to do so, we make use of a very powerful tool, Wilson's Renormalisation Group (RG) approach \cite{wilson_1, wilson_2}, carried out in momentum space; the framework was developed in the 70's on field theories dealing effectively with systems exhibiting scale invariant behavior. 

The subject of criticality in DL systems has been vastly addressed, see e.g. \cite{dl_crit}. The connection with the RG was proposed in \cite{rg_map_dl}, and implemented via block spin renormalisation e.g. in \cite{rg_block_bialek}. To our knowledge this is the first attempt to act with RG on the theory in momentum space.

The article is organised as follows: in section \ref{rg_primer} we present a high-view, intuitive summary of the RG concept, dealing with the transition between different scales and emergent properties of the system; in section \ref{sect:effect} the connection between the DL system  and a genuine field theory is made; here we map the fully connected graph to an effective theory of fields encoded in the Hamiltonian density;
in the subsequent section we formulate the RG in momentum space and act with the group on the effective field theory; this causes the Hamiltonian to "flow", tracing a path in the coupling space, along which the couplings themselves change; the latter change in the couplings is encoded in general differential equations as presented in section  \ref{sec:rg_eqs}, which will then be translated into constraining conditions for the connection weights of the system at hand in section \ref{sec:flow}. A simple measure for criticality, the 2-point correlation function is presented in section \ref{sect:corr_func}, which we can compute exactly for the Gaussian system; it then serves as a tool to probe the DL architecture at criticality. After addressing the whole theoretical setup, we implement the criticality constraints in section \ref{sect:exp_results}. We conclude this article in section  \ref{sect:outlook} and hint at future work.

\section{The Renormalisation Group technique}
\label{sect:rg_tech}

\subsection{An RG primer}
\label{rg_primer}

The Renormalization Group (RG) technique has its origin around problems dealing with scaling in effective field theories; as such, it is universally useful whenever the problem at hand shows scale invariance. In our particular case, the fluctuations (and with them the correlation) of a field $\phi$ are self-similar at various scales when the system is located at a special locus in the space of coupling parameters - called criticality.
Hence we can make use of the self-similarity of the system and implement a renormalisation scheme, the Wilson RG \cite{wilson_1, wilson_2}, which will produce equations and from them  consistency constraints on couplings of the system.

Pictorially, the problem at hand and its solution can be understood through an analogy to dynamical equations and their fractal behavior \cite{frac_rg}; given a real (iterative) map $\mathcal{M}_\mu : \mathbb{R} \rightarrow \mathbb{R}$, depending on some one-dimensional parameter $\mu\in\mathbb{R}$, we contemplate its image in $\mathbb{R}$. By carefully tuning the parameter $\mu$, we can navigate between trivially converging solutions, multi-modal oscillations and self-similar behavior. At the critical value $\mu^*$,
the map being scale invariant, its image will resemble some fractal,  in our case some one-dimensional fractal curve; at a given scale, we can identify a (small) recurrent pattern, of given size $b$; this is the fractal motive of the image, repeating itself at different scales. Zooming out from our starting point, we will change scale and also change resolution (by rescaling all lengths with $b$) at the new scale, in order to be able to compare present picture and recover the pattern we had previously discovered; 

To achieve that, conceptually the steps to be  implemented are as follows:

\renewcommand\labelenumi{(\theenumi)}
\begin{enumerate}
\item assume system displays scale invariance

this is ultimately what we want to achieve by carefully choosing our couplings 
\item probe system (Partition function) at slightly different scale

zoom out by a factor of b to search for the "pattern" at new scale
\item impose structural equality of Hamiltonian, cf. (1)
\item absorb changes and renormalisation of fields into couplings

change resolution at new scale to ensure comparability to starting point
\item solve for the fixed points of the mapping, which determines criticality
\end{enumerate}

We regard our system as a scale-dependent effective action functional - the partition function, encoded in the functional integral over the Hamiltonian $H$; the latter will depend on fields and couplings $r,g,u,\cdots$. During renormalisation, the RG will act on $H(r,g,u)$ as

\begin{align}
\label{h_flow}
\mathcal{R}_b H(r,g,u) = H'(r',g',u')
\end{align}

\begin{figure}[t!]
\centering
\includegraphics[width=8cm]{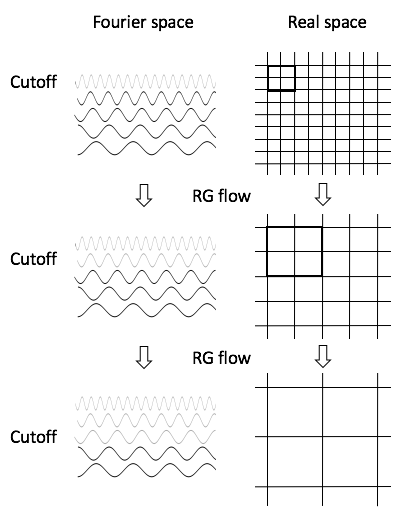}
\caption{Block-spin momentum space analogy, reproduced from K. Huang, Statistical Mechanics}
\label{fig:mom_space_cutoff}
\end{figure}

where $\mathcal{R}_b$ denotes the renormalisation operator and $b$ is the scaling parameter. Eq. (\ref{h_flow}) effectively describes the trace in the space of the couplings which is generated by the flowing Hamiltonian. Step (5) above singles out the special point in the couplings' space where

\begin{align}
H^* = \mathcal{R}_b H^*
\end{align}

holds, effectively meaning the system is invariant under scale change.

Moving now to practice, the program described above is implement the following way:

\subsubsection*{Coarse grain}
As depicted in fig. \ref{fig:mom_space_cutoff} we represent our system as a collection of units in the 2d place interacting with each other via couplings; in the configuration space we group units together into adjacent blocks of say $2^d$ units (2 per each site of the block) and consider their properties as a stand-alone unit:

\begin{equation}
\bar{\phi}(x) = \frac{1}{b^d}\int_{block} d^dy \phi(\mathbf{y})
\end{equation}

as depicted on the left in fig. \ref{fig:mom_space_cutoff}.
This step will reduce degrees of freedom from $N$ to $N'=N/4$, thus $b^d=\frac{N}{N'}=4$.

\begin{figure}[t!]
\centering
\includegraphics[height=6cm]{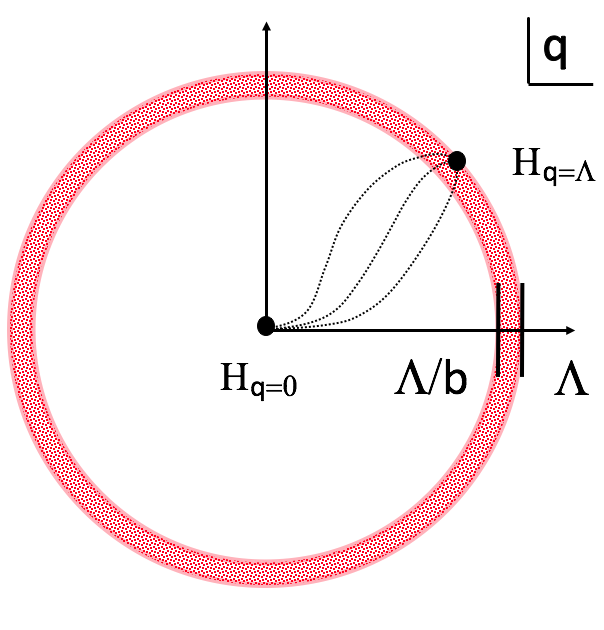}
\caption{Different functional paths of the Hamiltonian with shell of integration, between cutoff and renormalised momentum}
\label{fig:mom_shell}
\end{figure}

In the Wilsonian picture (which is usually the one employed in practical computations), all calculations are performed in momentum space; the calculations and the pictures are completely dual, however slightly more involved in the momentum picture; the relabeling of units into groups then corresponds to integrating out highest wave numbers $\mathbf{q}_>$; the high wave number obviously produce the highest resolution in the system and hence correspond to the smallest parts of the system - the very units; integrating out those wave numbers will be performed within the shell between $\Lambda/b \leq \mathbf{q}\leq \Lambda$, the latter being the natural cutoff of our effective field theory, see fig. \ref{fig:mom_shell}

\subsubsection*{Rescale}
In the newly regrouped picture we restore the original resolution by blowing up all lengths to the original scale

\begin{equation}
x' = \frac{x}{b}
\end{equation}

or in momentum space

\begin{equation}
\mathbf{q}' = b\mathbf{q}
\end{equation}

This is the equivalent of step (2) above, where we effectively "zoom out" by adjusting lengths to be comparable with original scale;

\subsubsection*{Renormalise}
After rescaling, the newly produced Hamiltonian is required to match in structure the starting one; all scaling factors from fields will be absorbed into couplings which effectively causes them to shift, or "flow". 
Ultimately this produces equations for them, whose fixed point solution will single out the fully scale invariant Hamiltonian.

\subsection{Effective field theory}
\label{sect:effect}

The Wilson RG works in the framework of effective field theories; in this chapter we will map our feed forward network setup to an effective field theory.
Starting point is the mean field consistency equation for a ReLU network as it has been computed in \cite{crit_dl1}:

\begin{align}
\label{mf_consi}
V_i &= \tanh\beta
(\textstyle\sum_{k}w_{ik}V_k/N + h)
\end{align}

Here $V_k$ are mean field variables, which are polynomials in $w_{ik}$ and $h$, once the consistency relations are solved; eq. (\ref{mf_consi}) stems from the stationarity condition imposed on the Hamiltonian

\begin{align}
\label{mf_ham}
H_V = &\frac{\beta}{2N}\textstyle\sum_{ij} w_{ij}V_iV_j \\
-&\sum_i\ln{\cosh \beta(\textstyle\sum_{j}w_{ij}V_j/N + h) } \nonumber
\end{align}

which is summed over all its states with the full partition function

\begin{align}
\boldsymbol{\mathcal{Z}} = c\prod_{k=1}^N\int dV_k\, e^{-H_V(w,h)}
\end{align}

The mean field equation will blow up at criticality, for $h\rightarrow 0$ and specific values $w_{ik}$, while the temperature approaches $T\rightarrow T_c$. 
Given the non-linear differential nature of the coupled consistency equations and the many limits which need to be taken, a solution in this case is rather cumbersome to obtain;

The RG technique however is designed to probe the system exactly at criticality, operating on a genuine field Hamiltonian. As shown in detail in appendix \ref{app1} we lift our effective variables to a genuine field theory, by effectively promoting  variables to fields (densities):

\begin{align}
\label{field_lift}
V_k &\rightarrow \phi_k(t) \\
\int dV_k &\rightarrow \int D\phi_k(t)\nonumber\\
H_V &\rightarrow \int d^dx H_\phi(w,h) \nonumber
\end{align}

The field functions depend on $d$-dimensional  spacial coordinates $\mathbf{x}$; the integral of the partition function then morphs into a functional integral 

\begin{align}
\label{eff_part}
\boldsymbol{\mathcal{Z}} = c\prod_{k=1}^N\int D\phi_k\, e^{-H_\phi(w,h)}
\end{align}

with the effective Hamiltonian

\begin{align}
\label{eff_h}
H_\phi =& \int d^dx\, \frac{\beta}{2N}\sum_{kl} [  w_{kl}\phi_k\phi_l
+ \delta_{kl}(\nabla_x\phi_k)(\nabla_x\phi_l) ]\nonumber\\
-&\sum_k\ln\cosh \beta(\textstyle\sum_{l}w_{kl}\phi_l/N + h)\, ]
\end{align}

\subsection{The action of RG}
\label{sect:rg_dl}

As explained in section \ref{rg_primer} the RG transformation $\mathcal{R}_b$ will map the Hamiltonian (\ref{eff_h}) structurally onto itself while scaling the parameters, hence tracing out the flow of the Hamiltonian in space spanned by the coupling constants.

Obviously we have strong non-linearities in our Hamiltonian, which need to be treated perturbatively; taking only the leading contributions from the $\ln\cosh$-term, we obtain the Gaussian model (in vectorial, coordinate-free notation):

\begin{align}
\label{gauss_x}
H_\phi = \int d^dx\,
[\, \frac{1}{2}\textbf{r}\cdot\boldsymbol{\phi}\cdot\boldsymbol{\phi}
+ \frac{1}{2}\mathbf{g}\cdot\boldsymbol{\phi}_x\cdot\boldsymbol{\phi}_x  - \mathbf{u}\cdot\boldsymbol{\phi}\, ] 
\end{align}

Here we have defined 

\begin{align}
\label{rgu}
\mathbf{r}&\equiv r_{kl} =\textstyle\frac{\beta}{N}(w_{kl}-\frac{\beta}{N}w^2_{kl}) , \\\nonumber
\mathbf{g} & \equiv g_{kl} =\textstyle\frac{\beta}{N}\delta_{kl} ,\\\nonumber 
\mathbf{u}&\equiv u_k=\textstyle\frac{\beta^2}{N}h\sum_l w_{kl},\\\nonumber
\boldsymbol{\phi}_x & \equiv \nabla_x\phi_k
\end{align}

and dropped the constant term $h^2$. 
Just for the sake of clarity, we have defined the bold constants $\mathbf{r}, \mathbf{g}, \mathbf{u}$ coordinate free; they are understood as (bi-)linear operators $\mathbf{o}\equiv\mathbf{o}(\_,\_)$ which take in vectors (in our case $\boldsymbol{\phi}$) and produce a scalar. Obviously from eq. (\ref{eff_h}) we know we have a collection of $N$ fields $\phi_i$ which interact via non-constant weights $w_{ij}$. We will use this operator, coordinate free language during our derivation for the RG equations, and only adopt coordinate notation, once we go to the component level.

The Gaussian model will be solved via expanding the functions $\boldsymbol{\phi}(x)$ wrt. a suitable base such that the Hamiltonian (\ref{gauss_x}) will be diagonalised; as explained in appendix \ref{app2}, the base turns out to be $\exp( i\mathbf{k}\mathbf{x})$, i.e. the Fourier basis.

Introducing the Fourier transformed fields 
\begin{align}
\boldsymbol{\phi(x)} &= \int\frac{d^dq}{(2\pi)^d}\,\boldsymbol{\phi(q)}e^{iqx} \\
\boldsymbol{\phi(q)} &= \int d^dx\,\boldsymbol{\phi(t)}e^{-iqx} \nonumber
\end{align}

and moving into momentum space we obtain (see eq. \ref{H_fourier})

\begin{equation}
\label{gauss_q}
H_\phi = \int\frac{d^dq}{(2\pi)^d} \frac{1}{2} 
(\mathbf{r} + \mathbf{g}q^2)\boldsymbol{\phi(q)}\cdot\boldsymbol{\phi(-q)}
- \mathbf{u}\cdot\boldsymbol{\phi(q=0)}
\end{equation}

We proceed now with the main three steps of the RG process as explained in section \ref{rg_primer}

\subsubsection*{Coarse grain}
We choose a coarse graining resolution $b$ via which we define the UV momentum region to be integrated out, as 
$\Lambda/b < |\mathbf{k}| < \Lambda$, and we separate the fields into high/low momentum regions

\begin{align}     
  \mathbf{m(q)} =\begin{cases}
   \mathbf{m_<(q)}, \, 0 < |\mathbf{q} < \Lambda/b\\
   \mathbf{m_>(q)}, \, \Lambda/b < |\mathbf{q} < \Lambda/b\\
            \end{cases}
\end{align}

With that, partition function takes the form

\begin{align}
\label{ir_uv_form}
\boldsymbol{\mathcal{Z}} = 
\int D\mathbf{m_<(q)} \int D\mathbf{m_>(q)}\,
e^{-\beta H[\mathbf{m_<}, \mathbf{m_>}]}
\end{align}

The low/high frequency fields decouple nicely in the Hamiltonian in (\ref{ir_uv_form}) and hence the high-frequency part can be integrated out to:

\begin{align}
&\boldsymbol{\mathcal{Z}} = \boldsymbol{\mathcal{Z_>}}
\int D\mathbf{m_<(q)}\\
&\exp\left[
-\int_0^{\Lambda/b} \frac{d^d\mathbf{q}}{(2\pi)^d}
\frac{r+g\mathbf{q}^2}{2}\mathbf{m_<(q)}^2 + u\mathbf{m_<(0)}
\right]\nonumber
\end{align}

while 
$\boldsymbol{\mathcal{Z_>}} = \exp[-L\textstyle\int_{\Lambda/b}^\Lambda \frac{d\boldsymbol{q}}{(2\pi)^d}\ln(\mathbf{r}+\mathbf{g}\boldsymbol{q}^2)] $, where $L$ is a numerical constant related to the volume of integration.

\subsubsection*{Rescale}

The integral for $\boldsymbol{\mathcal{Z_<}}$, representing the bulk of the modes, is now almost identical to the original one in the partition function $\boldsymbol{\mathcal{Z}}$, except for the upper limit of integration; by rescaling 
$\mathbf{q} \rightarrow \mathbf{q}'=b\mathbf{q}$ we restore the original cutoff $\Lambda$; however this will result in rescaling all quantities dependent on $\mathbf{q}$;

\subsubsection*{Renormalise}

This is the final step in the program, which renormalises the fields, aka the order parameter via 
$\boldsymbol{m(x')} \rightarrow \boldsymbol{m'(x')} = \boldsymbol{m_<(x')}/z$; from a pictorial point of view this will restore the resolution such that we can compare quantities from this scale with the quantities before scaling;

The partition reads now

\begin{align}
&\boldsymbol{\mathcal{Z}} = \boldsymbol{\mathcal{Z_>}}
\int D\mathbf{m'(q')}e^{-\beta H'[\boldsymbol{m'(q')}]}
\end{align}

with the term in the exponential $\beta H'$ given by

\begin{align}
\exp\left[
-\int_0^{\Lambda} \frac{d\mathbf{q}'}{(2\pi)^d}b^{-d}z^2
\frac{\boldsymbol{r}+\boldsymbol{g}b^{-2}\mathbf{q'}^2}{2}\mathbf{m'(q')}^2 + z\boldsymbol{u}\mathbf{m'}
\right]
\end{align}

At a glance we recognize that the singular point  $(\mathbf{r},\mathbf{u}=\{0,0\})$ is already a solution for the stationarity of the couplings; hence, in order to make the system scale invariant for this specific case,
we use the degree of freedom of our renormalisation to keep $\mathbf{g}=\mathbf{g}'$, which implies $z = b^{1+d/2}$. 

\subsection{Flow of the coupling constants}
\label{sec:rg_eqs}

By having fully determined the renormalisation freedom we obtain now the recursion relations for the couplings

\begin{align}
\label{couplings_rg}
\mathbf{r}' &= b^2\, \mathbf{r} \\
\mathbf{u}' &= b^{(d+2)/2}\, \mathbf{u} \nonumber
\end{align}

Assuming our first coarse graining step small, i.e. $b\approx1$,
we linearise and expand to first order
\begin{align}
b^n = (1+d\tau)^n \approx 1+nd\tau
\end{align}

and we obtain running equations of couplings for our system

\begin{align}
\label{running}
\frac{d\mathbf{r}}{d\tau} &= 2\mathbf{r}\\
\frac{d\mathbf{u}}{d\tau} &= \frac{d+2}{2} \mathbf{u} \nonumber\\
\frac{d\mathbf{g}}{d\tau} &= 0\nonumber
\end{align}

However, we remember the original coupling constants $w_{ij}, h$ relate to $\mathbf{r},\mathbf{g},\mathbf{u}$ via eq. (\ref{rgu}). Combining (\ref{running}), (\ref{rgu}) and going coordinate free, we finally obtain the famous running equations of the couplings

\begin{subequations}
\begin{align}
\frac{d}{d\tau} (\mathbf{w}-\frac{\beta}{N}\mathbf{w}\mathbf{w})
&= 2 (\mathbf{w}-\frac{\beta}{N}\mathbf{w}\mathbf{w})\label{beta_eq_w}\\
\frac{d}{d\tau} h\textstyle\mathbf{w}\mathbf{1}_v &= \frac{d+2}{2} h\mathbf{w}\mathbf{1}_v\label{beta_eq_h}
\end{align}
\end{subequations}

where we have introduced the linear operator

\begin{align}
\mathbf{w}\mathbf{1}_v &= \textstyle (\sum_l w_{kl})
\end{align}

which is the contraction of $w_{ik}$ with the one vector, $\mathbf{1}_v = (1,\hdots,1)$, and hence effectively summing over the contracted index.

We carry out the differentiation, denote $d\mathbf{w}/d\tau = \mathbf{w}_\tau$ and solve for $\mathbf{w}_\tau$ in (\ref{beta_eq_w}), after which we solve for $h_\tau$ in (\ref{beta_eq_h}), and simplify to

\begin{align}
\label{beta_matrix}
\mathbf{w}_\tau &= 
2 \mathbf{w}(\mathbf{1}- \mathbf{w}\beta/N)\left[\mathbf{1} -2 \mathbf{w}\beta/N \right]^{-1}\\
 h_\tau &= \frac{d+2}{2} h - h(\mathbf{w}\mathbf{1}_v)^{-1}\mathbf{w}_\tau \mathbf{1}_v 
\nonumber
\end{align}

The exponent $-1$ denotes the inverse of the linear operator, defined s.t. $\mathbf{O}^{-1}\mathbf{O}=\mathbf{O}\mathbf{O}^{-1}=\mathbf{1}_{N\times N}$.

The analysis of (\ref{beta_matrix}) is the topic of our next section.

\subsection{Constraining equations}
\label{sec:flow}

As explained in section \ref{rg_primer} we search for the point in parameter space, where couplings do not "run" anymore with the scaling, which mathematically translates into their derivatives (wrt. scaling parameter $\tau$) being zero

\begin{equation}
\partial_\tau
\begin{pmatrix} 
\mathbf{w} \\ \mathbf{h}  
 \end{pmatrix} \overset{!}{=} 0
\end{equation}

Just a glance at eq. (\ref{beta_matrix}) reveals already some first solutions. We will classify now all solutions in terms of their physical meaning and single out the critical point.

The equation for $\mathbf{w}_\tau$ in (\ref{beta_matrix}) does not depend on $h$, and hence can be solved on its own. Since we require $\mathbf{w}_\tau=0$, the solution of the second equation for $h_\tau$ in (\ref{beta_matrix}) also requires $h$ to be zero. We are thus left with classifying the solutions which lead to $\mathbf{w}_\tau=0$, and hence following cases: 

\begin{itemize}
  \item $\mathbf{w}=\mathbf{0}$
  \item $\mathbf{w}(\mathbf{1}- \mathbf{w}\beta/N)=\mathbf{0}$ 
  \item $(\mathbf{1}- \mathbf{w}\beta/N)=\mathbf{0}$ 
  \item $(\mathbf{1}- \mathbf{w}\beta/N)[\mathbf{1} -2 \mathbf{w}\beta/N ]^{-1}=\mathbf{0}$ 
  \item $\mathbf{w}(\mathbf{1}- \mathbf{w}\beta/N)[\mathbf{1} -2 \mathbf{w}\beta/N ]^{-1}=\mathbf{0}$ 
\end{itemize}

The case $[\mathbf{1} -2 \mathbf{w}\beta/N ]^{-1}=0$ cannot happen, as $\mathbf{0}$ cannot be the inverse matrix. Also we understand that in the cases involving products contain strict non-zero terms, only the product itself is zero.

\subsubsection*{Trivial solution, $T\rightarrow\infty$ }
First case represents the trivial solution  $(\mathbf{w}, h) = (0,0)$; this basically means $T\rightarrow\infty$, and zero correlation due to total disorder.

\subsubsection*{Trivial solution, $T\rightarrow 0$ }
The second bullet, can be written as
$(\mathbf{w}N- \mathbf{w}^2/T)=\mathbf{0}$. If we assume $\mathbf{w}$ to be of order $1/N$, then $\mathbf{w}^2$ is of order $1/N^2$ and the temperature has to cancel that term, effectively leading to $T=1/N^2\rightarrow 0$ in the large $N$ limit. Here we deal with perfect correlation, all units parallel, either up or down.
(We neglect the idempotent case $\mathbf{w}=\mathbf{w}^2$, since then $\mathbf{w}$ is either unity or singular.)

\subsubsection*{Critical CW system, $\mathbf{w}= c\mathbf{1}$}
Third bullet implies

\begin{align}
\label{cw_crit}
\frac{N\mathbf{1}}{\beta} &= \mathbf{w}\\
&\Updownarrow\nonumber\\
\frac{\mathbf{w}}{N} = const &= T\mathbf{1} \equiv T_c\mathbf{1}\nonumber
\end{align}

Eq. (\ref{cw_crit}) resembles the constant coupling case, $w_{ik}=J$ of a classical fully connected system which reaches criticality at a temperature $T_c=J$ when $h=0$, as discussed e.g. in \cite{cw_full_solution}

\subsubsection*{Critical, non-constant coupling}

The fourth  case reads $(\mathbf{1}- \mathbf{w}\beta/N)[\mathbf{1} -2 \mathbf{w}\beta/N ]^{-1}=\mathbf{0}$. We can expand the inverse term into its Neumann series 

\begin{equation}
[\mathbf{1} -2 \mathbf{w}\beta/N ]^{-1} = \sum_k (2\beta\mathbf{w}/n)^k    
\end{equation}

up to quadratic order and then obtain

\begin{align}
\label{crit_non_const}
&(\mathbf{1}- \mathbf{w}\beta/N)[\mathbf{1} -2 \mathbf{w}\beta/N ]^{-1} \\
\approx &(\mathbf{1}- \mathbf{w}\beta/N)[\mathbf{1}  +2\mathbf{w}\beta/N + 4\mathbf{w}^2(\beta/N)^2]\nonumber\\
= &(\mathbf{1} + \mathbf{w}\beta/N+ 2\mathbf{w}^2(\beta/N)^2)\nonumber
\end{align}

\subsection{Correlation function and scale invariance}
\label{sect:corr_func}

As shown in Appendix \ref{app3} we are able to fully solve our linear model and hence compute the 2-point correlation function for two nodes $k$ and $l$ and its power law behavior turns out to be

\begin{align}
\label{2_node_corr}
C_{kl} \sim \frac{1}{|\mathbf{x}|^{d-2}}
\end{align}

Eq. (\ref{2_node_corr}) shows the divergent (log) behavior of the function at criticality, i.e. the system exhibits scale invariance for the right choice of couplings and noise (temperature): $\mathbf{if}$ we can constrain the weight matrix $w_{ij}$ to obey the criticality conditions, $\mathbf{then}$ our system displays scale invariance through the power-law shaped correlation function.

This measure is a very handy tool to probe our real deep learning setup for long-range correlations, once we impose the fixed-points constraints (\ref{crit_non_const}). We can sample the node activations during the prediction epochs and hence register their activity and decide what kind of law they obey. As it turns out, once we impose the critical regularisation on our deep learning architecture, the node activation patterns will obey strong linear behavior on the log-log scale and hence display a power law behavior supporting the scale invariance.

\section{Experimental results}
\label{sect:exp_results}

\begin{figure*}[t!]
\centering
    \begin{minipage}{.45\textwidth}
    \centering
    \includegraphics[height=6cm]{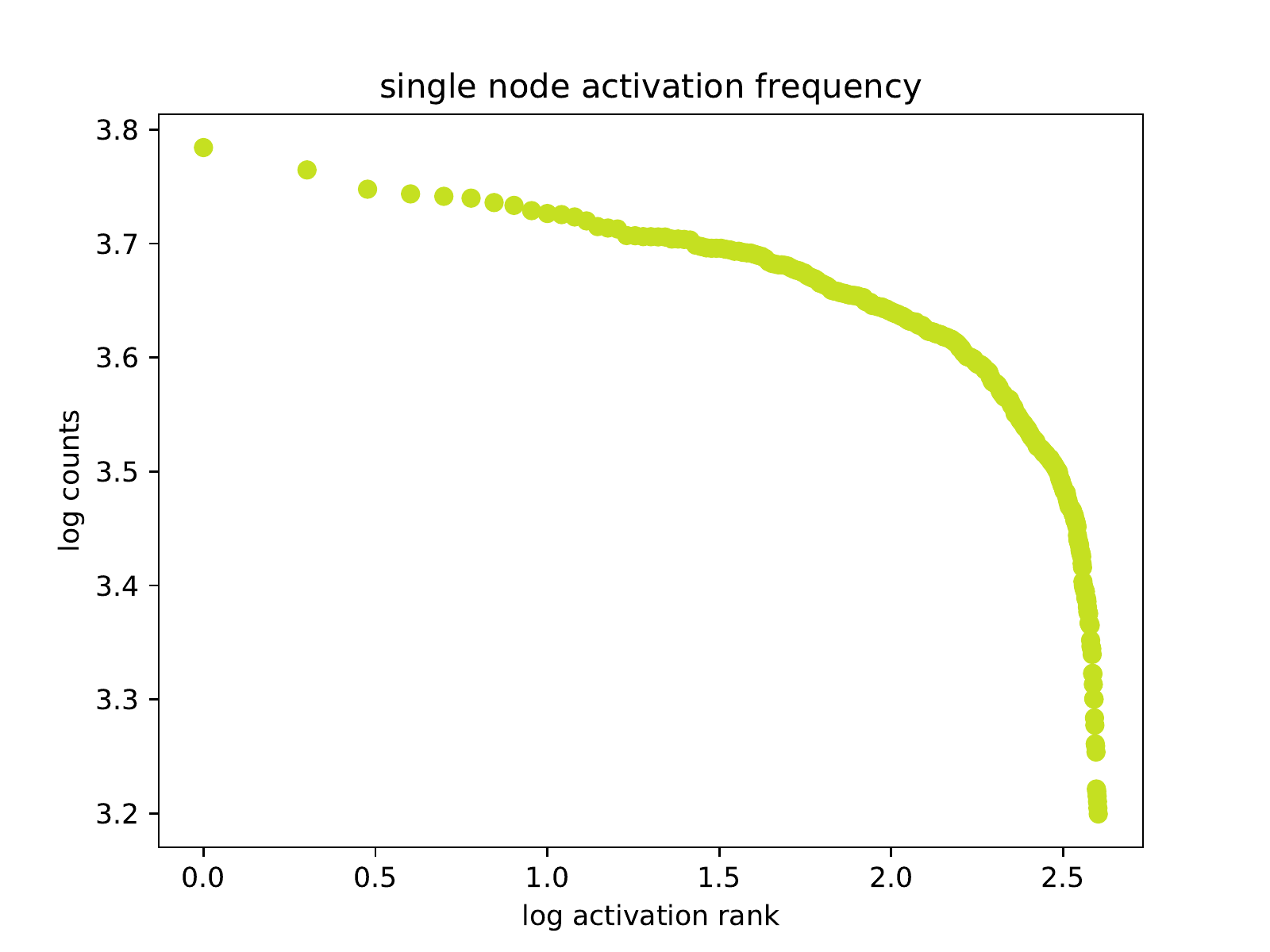}
    \caption{Activation ranks for Feed Forward network with no regularisation}
    \label{node-a}
    \end{minipage}\qquad
    \begin{minipage}{.45\textwidth}
    \centering
    \includegraphics[height=6cm]{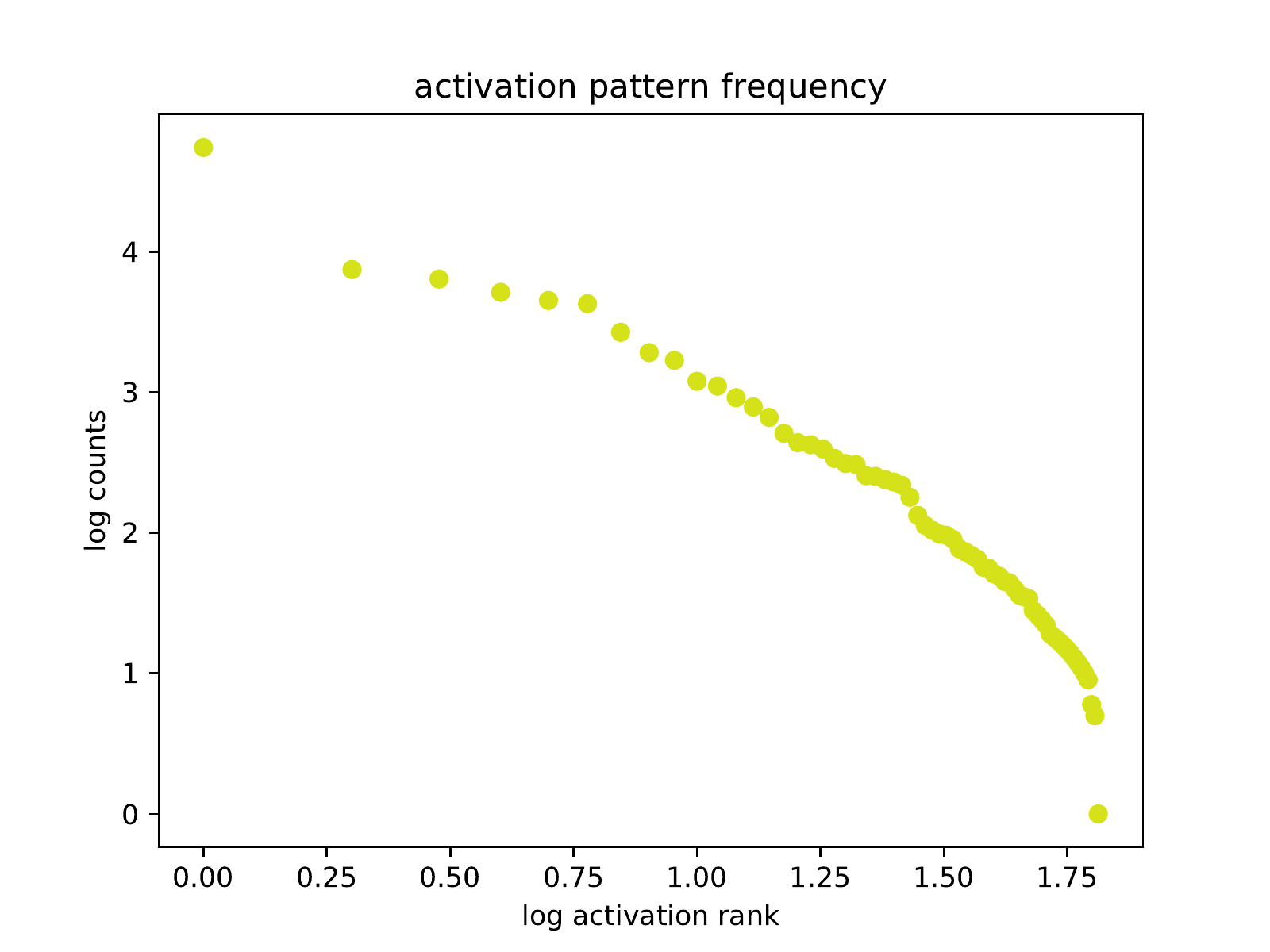}
    \caption{Activation ranks for Feed Forward network with critical regularisation}
    \label{node-b}
    \end{minipage}
\end{figure*}

We now move on to implementing the constraints found in section \ref{sec:flow}; as it turns out, a straightforward way of imposing those constraints on the system is modifying the loss function with an extra term containing the constraint; those constraints will hence translate into regularization terms, resembling elastic (L1, L2) regularization (and higher) given the linear and quadratic appearance of $\mathbf{w}$; 

Before tackling the constraining equations for $\mathbf{w}$ we have to address the fact, that $\mathbf{w}$ describes a fully connected system, which, to first order, resembles the weight matrix between two layers of a feed forward architecture, cf. appendix B of \cite{crit_dl1}:

\begin{equation}
\label{v_w}
\underbrace{\begin{bmatrix} v_{11} & \cdots & v_{1n} \\ \vdots & \ddots & \vdots \\ v_{m1} & \cdots & v_{mn} \end{bmatrix}}_{\mathbf{v}^{m\times n}}
\xymatrix{    
 \ar@/^0.4cm/[r]    &
}
\,\,\mathbf{v}\times\mathbf{v}^\intercal\equiv
\underbrace{\sum_k v_{ik}v_{kj}}_{\mathbf{w}^{m\times m}}
\end{equation}

On the left we have a bipartite graph representing a 2-layer feed forward system with $m$ respectively $n$ units connected via the weight matrix $\mathbf{v}$; this is equivalent to first order to a fully connected layer of $m$ units, while the connection matrix $\mathbf{w}$ is a function of the feed forward weight matrix as depicted in eq. (\ref{v_w}).

Starting from (\ref{crit_non_const}) we switch to coordinate language and obtain

\begin{align}
    \frac{\delta_{i,j}\beta w_{ij}}{N} + \sum_{k}w_{ik}w_{kj}\frac{2\beta^{2}}{N^{2}} = -\delta_{i,j}
\end{align}

Those are component-wise constraints on the weight matrix $w_{ij}$ which, when satisfied, will induce criticality and hence scale invariance in our system.

\subsection{Critical regularization}
\label{sect:crit_reg}

As computed in section \ref{sec:flow} we have two cases of interest where scale invariance will be induced:

\begin{itemize}
  \item constant $\mathbf{w}$, i.e.  $(\mathbf{1} = \mathbf{w}\beta/N)$ 
  
  \item $(\mathbf{1} + \mathbf{w}\beta/N+ 2\mathbf{w}^2(\beta/N)^2)$ 
\end{itemize}

While the first equation addresses the constant weight matrix, i.e. a multiple of unity, the second equation implements a non-trivial solution of criticality and hence it will be our case of study.

For our experiments we used the CIFAR- 10 dataset for all investigated models; furthermore, our architecture relies on ReLU/eLU activations while for the optimisation we use the Adam Optimizer without gradient clipping. 
We implemented a feed forward network with 4 layers, of 600, 400, 200 and 100 nodes respectively.
Our focus was mainly inducing scale invariance and exactly capturing the regime of its emergence;

\subsubsection*{Layer activation}

In figure \ref{node-a} we depict the activation ranks of a normal feed forward architecture without regularisation; for the layer activation patterns we counted the frequency of each layer's activation through the inference epochs and then we sorted those by rank; the figure then depicts the log counts versus the logged ranks. Next to it, we have implemented the critical regularization, in figure \ref{node-b}. We obtain a strong deviation from the non-regularized system: where on the left the system is almost linear and then abruptly falls off towards higher ranks, with critical regularization the activation is nearly linear and stays that way until the very end of the distribution; also the slope of the distribution is very steep, hence once more distinguishing it from the "normal" case; this strong linear behavior, implying a power law distribution is the prime indicator for scale invariance, as discussed in section \ref{sect:corr_func}.

\subsubsection*{Average node activation}

\begin{figure}[t]
    \centering
    \includegraphics[height=6cm]{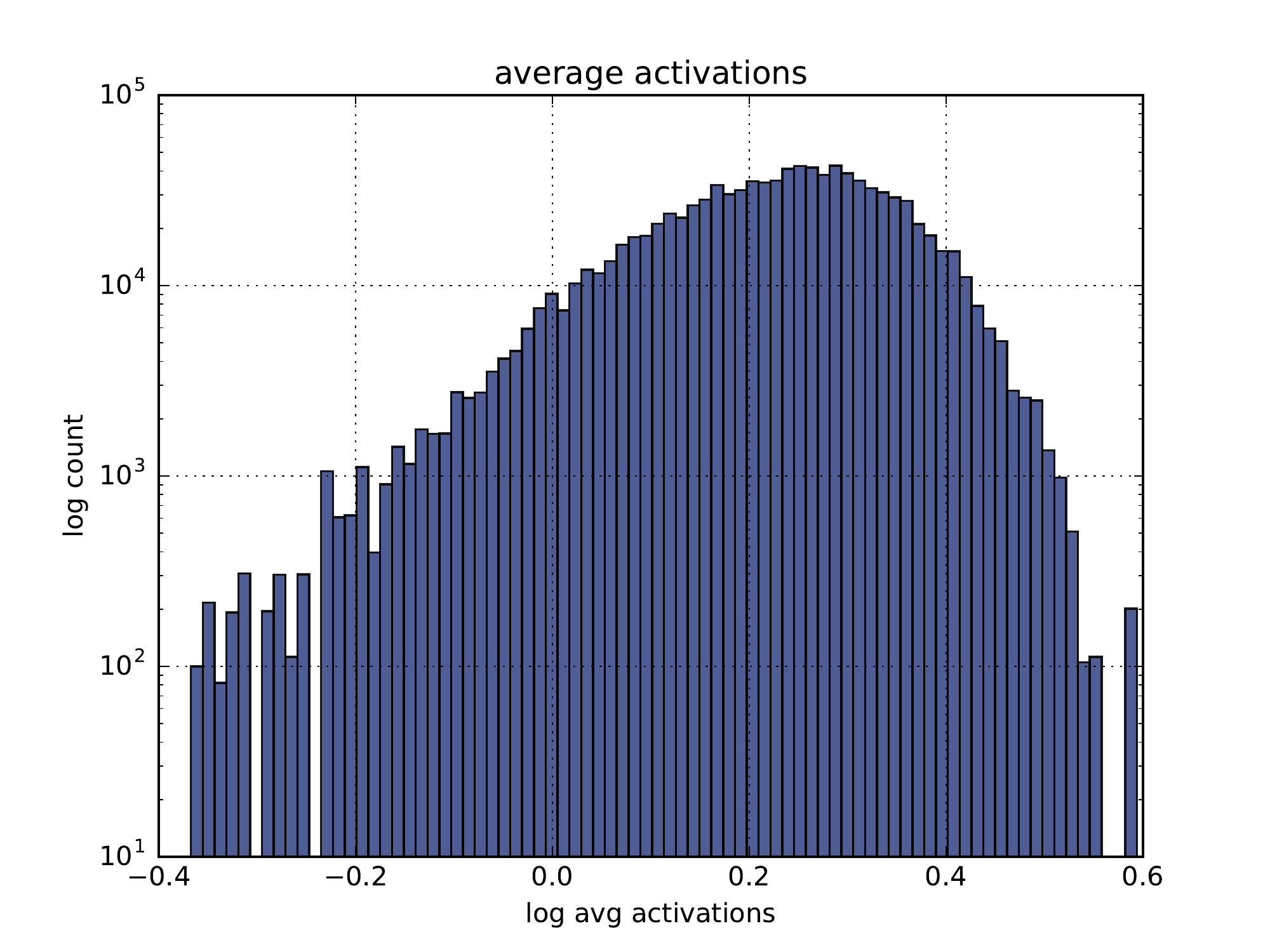}
    \caption{Ranks of average layer activations for Feed Forward network with no regularisation}
    \label{avg-a}
    \includegraphics[height=6cm]{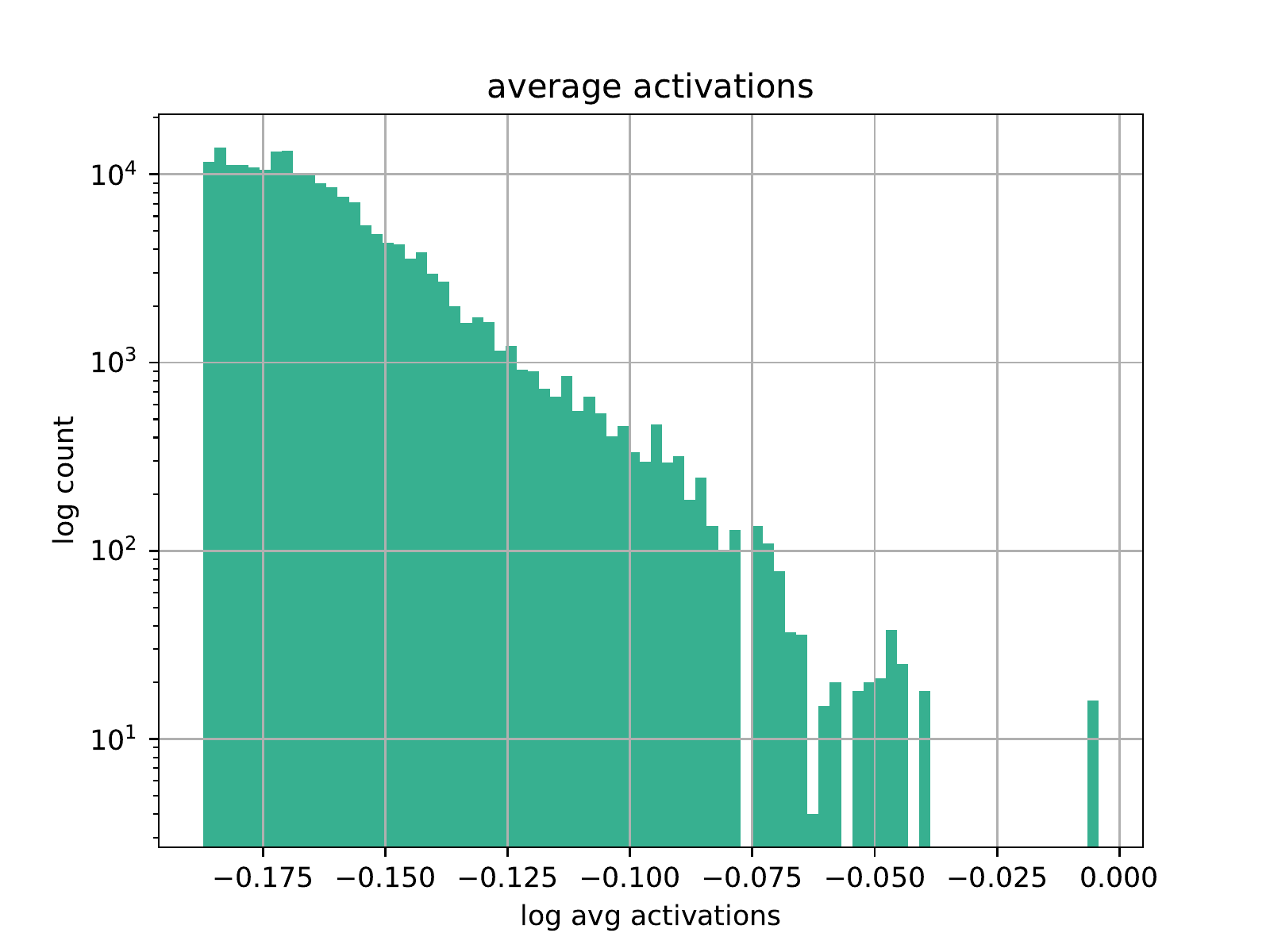}
    \caption{Ranks of average layer activations for Feed Forward network with critical regularisation}
    \label{avg-b}
\end{figure}

Another measure we employed in detecting deviating behavior in critically regularized systems is the average activation of the nodes during the prediction epochs. Given a layer, we averaged over the activations of all units in that layer for one prediction epoch, after which we ranked the log averages by their log counts - the results are visible in figure \ref{avg-a} and \ref{avg-b}. 

The top graph depicts the log-log distribution of 4-layered feed forward net with no regularization, contrasting to the graph below it, where the distribution comes from same architecture but with critical regularization employed. 
The linear behavior is strongly visible, over four orders of magnitude in the count of the ranks; hence another criterion supporting the scale invariance of the architecture tuned rightly via regularization.

\subsubsection*{Weighted degree distribution}

\begin{figure}[b!]
    \centering
    \includegraphics[height=6cm]{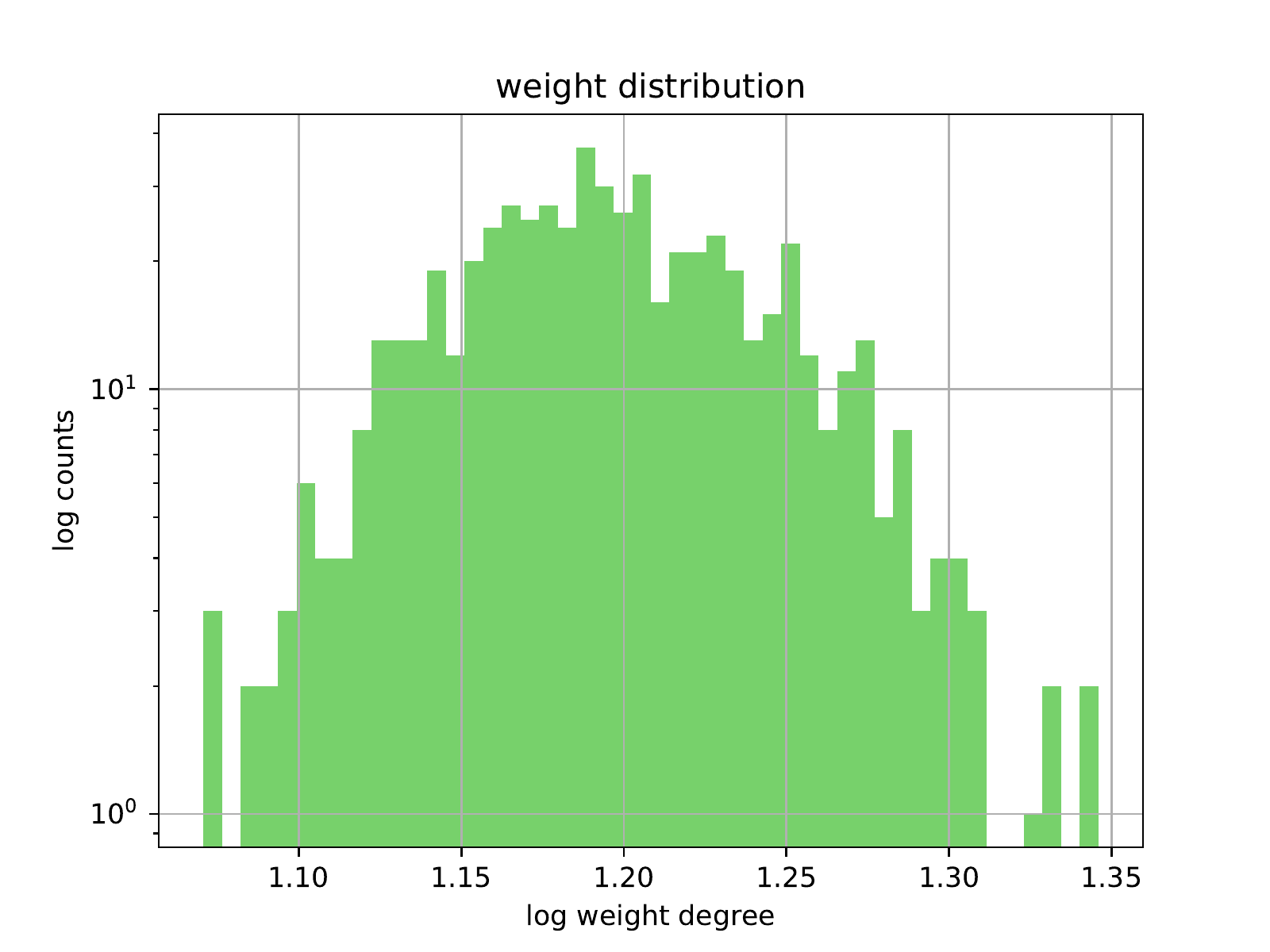}
    \caption{Log distribution of weighted degree of nodes per layer without regularisation}
    \label{deg-a}
    \includegraphics[height=6cm]{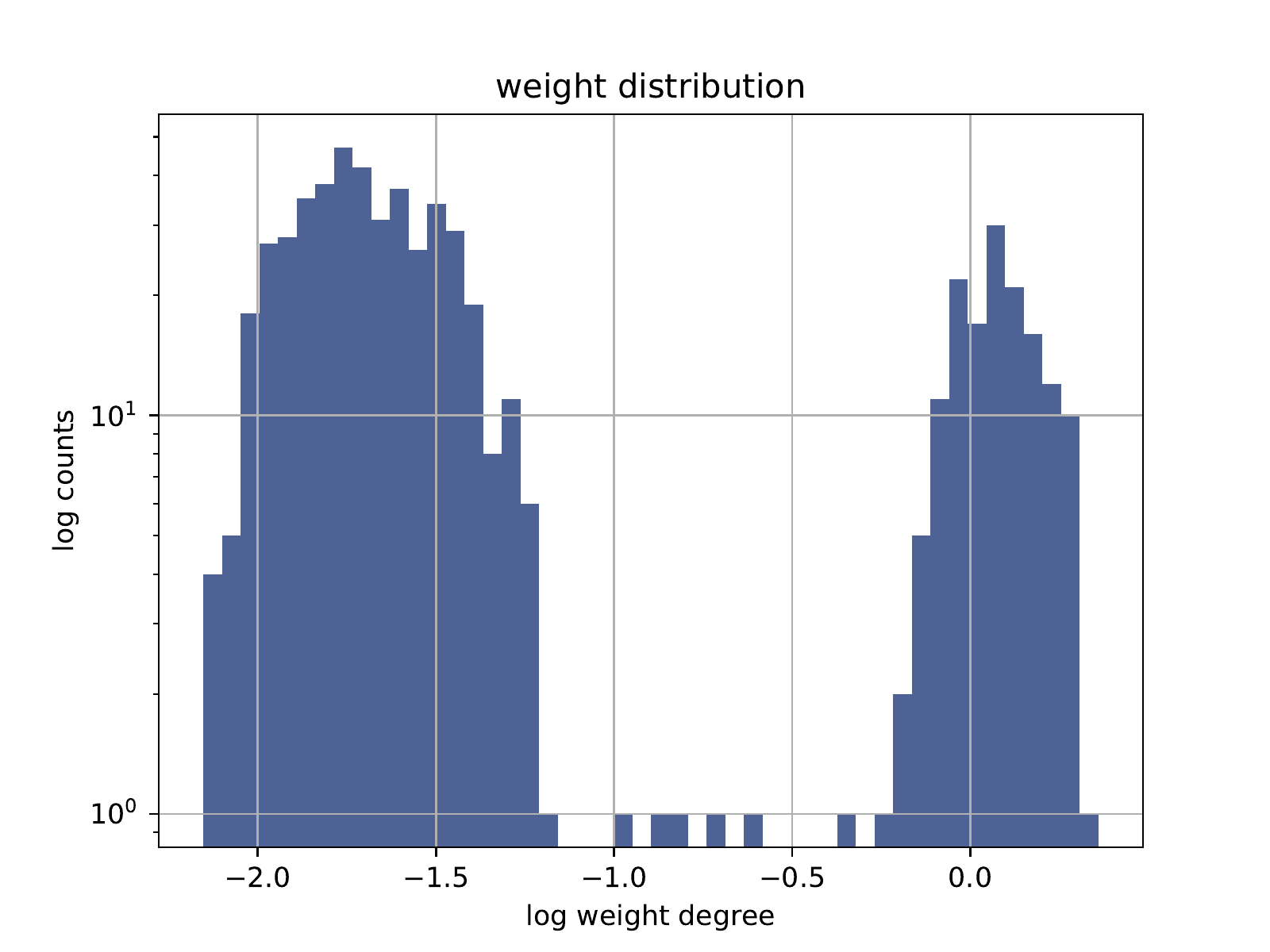}
    \caption{Log distribution of weighted degree of nodes per layer with critical regularisation}
    \label{deg-b}
\end{figure}

A last measure we used to test the validity of our mechanism is the weighted node degree, as suggested in \cite{complex_nets}. Here we sum all the values of the weights going out from a node; this is a weighted sum of the outgoing connections, as zero weights do not contribute and finite weights make a contribution weighted with unity. To every node in a layer we will hence attach the real value of its weighted degree; once again, we log-count the occurrences and plot against the logged degree, as depicted in picture \ref{deg-a} and \ref{deg-b}. The green graph depicts the degree distribution of the four-layer architecture without any regularization, while below it we have the same architecture subjected to critical regularization. The difference is quite dramatic, as the degree in the critical case exhibit a drastic bi-modal distribution, roughly around 1 and some other fractional value. Once again, we interpret this bi-modal distribution as the results of the inhomogeneous polynomial regularization employed.

\subsection{Applicability of our results}
\label{sect:unit_corr}
We conclude the experimental section stressing our approximations and shortcomings while arriving at the theoretical and experimental results depicted. 
All our calculations so far have been performed in a system where the units take on  values in $\{\pm1\}$; this was due to the analytic behavior of the results and hence the tractability of calculations. The domain of the feed-forward ReLU network though, is contained in $[0,+\infty]$; the translation from one domain into the other leaves the structure of the Hamiltonian unaltered and has as effect re-defined couplings; given the preserved structure of the Hamiltonian, the polynomial nature of the constraints will stay conserved, possibly with corrections in the coefficients; we regard thus the results of the RG transformation as a powerful hint towards non-homogeneous polynomial regularisation, which we have implemented above.

\section{Summary and outlook}
\label{sect:outlook}

Summary: By mapping a classical deep learning architecture to an effective theory of field (densities) we are able to employ the powerful tool of momentum space renormalisation for scale-free systems in the realm of deep learning networks.
Carrying out the renormalisation steps in momentum space we induce the flow of the coupling constants, while keeping the Hamiltonian structure unchanged; the flow of the constants are a set of non-linear differential equations which, when solved, employ strong conditions on the couplings and hence on the parameters of the deep learning system. The constraints are further translated into regularisation conditions, which take form of a non-homogeneous polynomial in the weight matrix.
We then implement this critical regularisation  and induce typical behavior of the net as observed in scale-invariant systems. In our experiments we use various metrics to measure the degree of scale invariance and detect clearly its presence.

Outlook: Despite the concreteness of the multy-layer feed forward network, we still lack accuracy in our mapping and neglect many details in our mapping, such as the values of the units and the multi-layer nature of the architecture. It would be of tremendous importance to address a full architecture, including its non-linearities in an analytical way. Ideally, the self-similarity of the network would be ported into deeply manifest group symmetries of the analytic counterpart.
This however, remains to be studied in future work.


\appendix

{\Large \bf  \centering{Appendix} \par}
\bottomtitlebar

\section{From variables to fields}
\label{app1}

In this section we will depict the steps in order to lift our classical fully connected layer system to an effective filed theory.

Given a functional $L(\phi(x))$ depending on (products) of the field $\phi(x)$ the functional (path) integral is defined as a formal infinite product of integrals over all the $x$:

\begin{align}
\label{path_formal}
\int D\phi(x) L(\phi) = \prod_x\int d\phi(x) L(\phi(x))
\end{align}

which in practice means discretising $x$ into $l$ supports $x\rightarrow x_k, k\in[-l,\hdots,+l]$, and then taking the limit

\begin{align}
\label{path_pract}
\prod_x\int d\phi(x) = \lim_{l\rightarrow\infty}\prod_{k\in[-l,\hdots,+l]}\int d\phi(x_k)
\end{align}

Further, we will generalize the Gaussian integral

\begin{align}
\int_{-\infty}^\infty \frac{dx}{\sqrt{2\pi}} e^{-(1/2)ax^2 + bx} = \frac{1}{\sqrt{a}}e^{b^2/(2a)}
\end{align}

to its functional version. Introducing the coordinate free notation $\langle\mathbf{a},\mathbf{b}\rangle=\textstyle\sum_ka_kb_k$ and
$\langle\mathbf{a},\mathbf{w}\mathbf{b}\rangle=\textstyle\sum_{kl}a_kw_{kl}b_k$ for the linear inner product, the generalization of the Gaussian integral to the functional case reads

\begin{align}
\label{funct_gauss}
\int D\phi\, &\exp[-\frac{1}{2}\langle{\boldsymbol{\phi}},\mathbf{K}\boldsymbol{\phi}\rangle +  \langle\boldsymbol{\eta},\boldsymbol{\phi}\rangle]\\
=& \frac{1}{\sqrt{\det{K}}}\, \exp\textstyle\frac{1}{2}\langle\boldsymbol{\eta},\mathbf{K}^{-1}\boldsymbol{\eta}\rangle\nonumber
\end{align}

We are now able, using these tools, to lift our system to an effective field theory; as explained in \cite{crit_dl1}, our Hamiltonian and associated partition function read in coordinate free notation

\begin{align}
\label{part_cw}
\boldsymbol{\mathcal{Z}} = \sum_{\mathbf{s}\in\{\pm1\}}e^{\textstyle-\frac{1}{2}
\langle\mathbf{s},\mathbf{w}\mathbf{s}\rangle - \langle\mathbf{h},\mathbf{s}\rangle}
\end{align}

with $\mathbf{s}=s_k$ being the N units, while $\mathbf{w}=w_{kl}$ being the fully connecting weight matrix. We insert now the relation (\ref{funct_gauss}) for the quadratic part $\exp[\textstyle-\frac{1}{2}\langle\mathbf{s},\mathbf{w}\mathbf{s}\rangle]$ of the partition function in (\ref{part_cw}) and obtain:

\begin{align}
\label{func_iden}
\boldsymbol{\mathcal{Z}}&={\textstyle\sum_{\mathbf{s}\in\{\pm1\}}}
\frac{1}{\sqrt{\det{\mathbf{w}}}}\\
&\int D\boldsymbol{\phi}
\exp[\textstyle-\frac{1}{2}\langle\boldsymbol{\phi},\mathbf{w}^{-1}\boldsymbol{\phi}\rangle + \langle\boldsymbol{\phi},\mathbf{s}\rangle]
e^{\langle\mathbf{h},\mathbf{s}\rangle}\nonumber\\
=&\frac{1}{\sqrt{\det{\mathbf{w}}}}\,\int D\boldsymbol{\phi}
\exp[\textstyle-\frac{1}{2}\langle\boldsymbol{\phi},\mathbf{w}^{-1}\boldsymbol{\phi}\rangle]{\textstyle\sum_{\mathbf{s}\in\{\pm1\}}}e^{\langle\mathbf{h}+\boldsymbol{\phi},\mathbf{s}\rangle}\nonumber\\
=&c \prod_k \int D\boldsymbol{\phi}_k
e^{-H(\phi_k,w,h)}\nonumber
\end{align}

while the Hamiltonian reads now

\begin{align}
\label{eff_hamil}
H(\phi,w,h) &= \\
&\textstyle\frac{1}{2}\langle\boldsymbol{\phi},\mathbf{w}^{-1}\boldsymbol{\phi}\rangle
\textstyle- \ln\sum_{\mathbf{s}\in\{\pm1\}}\langle\mathbf{h}+\boldsymbol{\phi},\mathbf{s}\rangle\nonumber
\end{align}

We have effectively restricted the sum over $\mathbf{s}\in\{\pm1\}$ in eq. (\ref{func_iden}) over the linear term only, by introducing the effective field $\mathbf{\phi}$. Hence we can now calculate the partition sum in eq. (\ref{eff_hamil}) and after one more
transformation $\textstyle\phi_k\rightarrow \sum_iw_{ik}\phi_i$, while neglecting the constant $N\ln2$  will bring us to the Hamiltonian:

\begin{align}
\label{field_hamil}
&H(\phi,w,h) = \\
&\frac{1}{2}\sum_{kl}w_{kl}\phi_k\phi_l -
\sum_k\ln\cosh(\sum_iw_{ik}\phi_i+{\phi_k})\nonumber
\end{align}

Last transformation will also produce a Jacobian equal to $\det\mathbf{w}$ multiplying the partition function.

We effectively traded the quadratic binary sum for additional fields $\phi_i$, while the remaining linear binary sum can be analytically computed;

In analogy to free energy per unit, we introduce the Hamiltonian density (per space) and hence think of the fields $\phi_k$ as density of the order parameter, which also display fluctuations (beyond the microsopic/atomic scale); the Hamiltonian density then reads:

\begin{align}
\label{hamil_density}
H(\phi,w,h) =& \int dx\, \frac{1}{2}\sum_{kl} [  w_{kl}\phi_k\phi_l
+ \delta_{kl}(\partial_x\phi_k)(\partial_x\phi_l) ]\nonumber\\
-&\sum_k\ln\cosh (\textstyle\sum_{l}w_{kl}\phi_l + h)
\end{align}

The $\phi_k$  are now genuine effective field functions, depending on spacial coordinates $\mathbf{x}$. The additional term containing the spacial derivative takes into account that  $\phi_i\equiv \phi_i(\mathbf{x})$, hence fields being dynamic and hence able to fluctuate, on larger scales than the next neighbour distance;

\section{Fourier transformed field theory}
\label{app2}

Eq. (\ref{hamil_density}) encodes all information of interest describing the system, which can be expressed in terms of correlation functions of various degree (i.e. the coordinated firing of $n$ random units through the architecture, as a function of their "distance", which is their index separation);

The term $\ln\cosh$ has to be treated perturbatively anyway, hence we will ignore it for now; the first part of (\ref{hamil_density}) is the "free" Hamiltonian, which can be fully diagonalised and solved, while the non-linear part can be expandede and treated as a correction term;

Integrating by parts the derivative term, we obtain a quadratic form

\begin{equation}
\label{kernel}
H(\phi,w,h) = -\int d^dx\,d^dy\, \sum_{kl} \phi_k \mathbf{M}_{kl}(\mathbf{x},\mathbf{y}) \phi_l
\end{equation}

with the operator $\mathbf{M}$ defined as  

\begin{equation}\mathbf{M}_{kl}(\mathbf{x},\mathbf{y})\equiv \frac{1}{2}\delta(\mathbf{x}-\mathbf{y})(w_{kl} - \delta_{kl}\nabla^2_x )
\end{equation}

By partial integration we picked up a crucial minus sign in front of the Laplace operator, which will prove very important in the solution of the system. Eq. (\ref{kernel}) can be fully diagonalised and hence solved once we find a suitable basis $\psi_q$, s.t. $\mathbf{M}$ acts linearly on it

\begin{equation}
\label{eigen_vec_val}
\mathbf{M}\boldsymbol{\psi}_q = \lambda_q\boldsymbol{\psi}_q
\end{equation}

We then expand our fields in the eigenvectors

\begin{equation}
\boldsymbol{\phi} = \sum_q\phi_q\boldsymbol{\psi}_q
\end{equation}

with $\phi_q$ given by the relation

\begin{equation}
\phi_q = \int dx \boldsymbol{\psi}_q^*(x)\boldsymbol{\phi}(x)
\end{equation}

The form of $\mathbf{M}$ dictates the choice for the eigenvectors

\begin{equation}
\label{eigen_vec}
\boldsymbol{\psi}_q = \exp(i\mathbf{q}\mathbf{x})
\end{equation}

Inserting (\ref{eigen_vec}) into (\ref{eigen_vec_val}) we obtain for the eigenvalues

\begin{equation}
\lambda_q = (\delta_{kl}q^2+w_{kl})/2
\end{equation}

The explicit expansion of the fields reads now

\begin{equation}
\boldsymbol{\phi}(x) = \int_{|q|<\Lambda}\frac{d^dq}{(2\pi)^d}\,\phi(q)e^{iqx}
\end{equation}

with $\phi(\mathbf{q})$ given by

\begin{equation}
\phi(\mathbf{q}) = \int d^dx\,e^{-i\mathbf{q}\mathbf{x}}\phi(\mathbf{x})
\end{equation}

and the Hamiltonian

\begin{align}
&H(\phi,w,h) = \\ 
&\int_{|\mathbf{q}|<\Lambda} \frac{d^dq}{(2\pi)^d}\,\frac{1}{2}\sum_{kl} (\delta_{kl}q^2+w_{kl})\phi_k(\mathbf{q})\phi_l(-\mathbf{q})\nonumber
\end{align}

where we have used the identity $\int d^x e^{i\mathbf{q}\mathbf{x}}e^{i\mathbf{p}\mathbf{x}}=(2\pi)^d\delta(\mathbf{q-p})$, which is the normality condition of the basis (\ref{eigen_vec}).

This is the diagonalised free Hamiltonian.

Given the effective nature of our theory, we have introduced a natural UV-cutoff $\Lambda$ in order to account for the finite validity of the Hamiltonian; 
in the partition function, also the integration measure will naturally change from paths in configuration space

\begin{equation}
\int D\boldsymbol{\phi}(\mathbf{x}) \rightarrow \int D\boldsymbol{\phi}(\mathbf{q})
\end{equation}

to paths over momenta once we transition to Fourier space.

Going now back to the original, full Hamiltonian and expanding the $\ln\cosh$-term to first order, grouping linear and quadratic terms together and going coordinate-free we finally obtain

\begin{equation}
\label{H_fourier}
H_\phi = \int\frac{d^dq}{(2\pi)^d} \frac{1}{2} 
(\mathbf{r} + \mathbf{g}q^2)\boldsymbol{\phi(q)}\cdot\boldsymbol{\phi(-q)}
- \mathbf{u}\cdot\boldsymbol{\phi(q=0)}
\end{equation}

with

\begin{align}
\label{bold_rgu}
\mathbf{r}&\equiv r_{kl} =\textstyle(w_{kl}-\sum_i w_{ki}w_{il}) , \\\nonumber
\mathbf{g} & \equiv g_{kl} =\delta_{kl} ,\\\nonumber
\mathbf{u}&\equiv u_k=\textstyle h\sum_l w_{kl}\nonumber
\end{align}

Obviously in the base (\ref{eigen_vec}) the derivative term produces only a multiplicative momentum factor.
The partition function based on the Gaussian Hamiltonian in momentum space reads:

\begin{align}
\label{gauss_pf}
&\boldsymbol{\mathcal{Z}} = c\int D\boldsymbol{\phi}(\mathbf{q}) \\
&e^{-\beta \textstyle  
\int\frac{d^dq}{(2\pi)^d} \frac{1}{2} 
(\mathbf{r} + \mathbf{g}q^2)\boldsymbol{\phi(q)}\cdot\boldsymbol{\phi(-q)}
- \mathbf{u}\cdot\boldsymbol{\phi(q=0)}
}\nonumber
\end{align}

The functional integral $D\boldsymbol{\phi}(q)$ is understood to be an infinite product  $D\boldsymbol{\phi}(q)=\prod_q\int d\boldsymbol{\phi}(q)$ over the momentum $q$, while each
$\boldsymbol{\phi}(q_k)$ is fixed at a specific location $q_k$. The constant $c$ multiplying the partition function contains the determinant and further numerical constants which only appear additive in the free energy $\boldsymbol{F}= -kT\ln{\boldsymbol{\mathcal{Z}}}$ and hence do not contribute anything neither to derivatives nor to normalised quantities, such as the correlation function.

\section{Solution of the Gaussian model}
\label{app3}

The functional integral (\ref{gauss_pf}) is a Gaussian type of integral and hence, luckily, can be fully solved; we arrived at it while lifting the theory to an effective field theory via (\ref{funct_gauss}); solving thus the Gaussian is simply reversing this very equation:

\begin{align}
\label{gen_funct}
\boldsymbol{\mathcal{Z}} = &\int D\boldsymbol{\phi}
e^{-\beta [\textstyle  \frac{1}{2} 
\langle\boldsymbol{\phi},\mathbf{K}\boldsymbol{\phi}\rangle - \langle\mathbf{u},\boldsymbol{\phi}\rangle
]}\\
=&\frac{1}{\sqrt{\det{K}}}\, \exp\textstyle\frac{1}{2}\langle\boldsymbol{u},\mathbf{K}^{-1}\boldsymbol{u}\rangle
\label{gen_corr}
\end{align}

where we have identified the operator
$\mathbf{K} = (\mathbf{r} + \mathbf{g}q^2)$ and introduced the inner product notation $\langle\mathbf{a},\mathbf{b}\rangle = \int\frac{d^dq}{(2\pi)^d}\mathbf{a}(\mathbf{q})\cdot\mathbf{b}(-\mathbf{q})$. The partition function $\boldsymbol{\mathcal{Z}}\equiv\boldsymbol{\mathcal{Z}}(\mathbf{u})$ in (\ref{gen_funct}) is also called the generating functional, for the obvious reason that we can generate from it n-point correlation functions; those are the average correlation functions for $n$ random units, as a function of their separation. Generally speaking, the average of an operator is given by

\begin{equation}
\langle\mathbf{O}\rangle \stackrel{\text{def}}{=} 
\frac{1}{\boldsymbol{\mathcal{Z}}(0)}\int D\boldsymbol{\phi} \mathbf{O}\,
e^{-\beta \textstyle  \frac{1}{2} 
\langle\boldsymbol{\phi},\mathbf{K}\boldsymbol{\phi}\rangle} \equiv \langle\mathbf{O}\rangle_0
\end{equation}

Here $\langle\mathbf{O}\rangle_0$ denotes the average  of operator $\mathbf{O}$ being taken wrt. $\boldsymbol{\mathcal{Z}}(0)\equiv\boldsymbol{\mathcal{Z}}(\mathbf{u}=0)$ 

Since we are interested mostly in the 2-point function, we will compute it here as: 

\begin{align}
&C_{kl}\equiv\langle\phi_k\phi_l\rangle_0=\\
&\frac{1}{\boldsymbol{\mathcal{Z}}(0)}
\int D\boldsymbol{\phi} \phi_k\phi_l\,
e^{-\beta \textstyle  \frac{1}{2} 
\langle\boldsymbol{\phi},\mathbf{K}\boldsymbol{\phi}\rangle}=
\nonumber\\
&\frac{1}{\boldsymbol{\mathcal{Z}}(0)}
\left.\int D\boldsymbol{\phi} \frac{\delta^2}{\delta u_k\delta u_l}\,
e^{-\beta [\textstyle  \frac{1}{2} 
\langle\boldsymbol{\phi},\mathbf{K}\boldsymbol{\phi}\rangle - \langle\mathbf{u},\boldsymbol{\phi}\rangle
]}  \right\rvert_{\mathbf{u}=0}
\nonumber\\
\,&=\left.\frac{\delta^2}{\delta u_k\delta u_l}
\ln\boldsymbol{\mathcal{Z}}(\mathbf{u}) \right\rvert_{\mathbf{u}=0}
\label{gen_der}
\end{align}

hence this justifies the name "generating functional" for $\boldsymbol{\mathcal{Z}}(\mathbf{u})$.

We can apply now (\ref{gen_der}) on (\ref{gen_corr}) to yield the explicit correlation function between two units

\begin{align}
\label{2_point}
C_{kl} = \langle\phi_k\phi_l\rangle_0 = \beta \int \frac{d^dq}{(2\pi)^d}\frac{e^{-i\mathbf{q}\mathbf{x}}}{\mathbf{r} + \mathbf{g}q^2}
\end{align}

We recall the definition of $\mathbf{r}, \mathbf{g}$ given in (\ref{bold_rgu}) and hence we recognize $\mathbf{K}^{-1}$ as a matrix inverse.

In order to get an impression of the form and especially of the asymptotic behavior of the correlation function (\ref{2_point}) we can rewrite it and proceed as follows:

\begin{align}
\label{2_point_sad}
&\int \frac{d^dq}{(2\pi)^d}\frac{e^{-i\mathbf{q}\mathbf{x}}}{\mathbf{r} + \mathbf{g}q^2} =
\int \frac{d^dq}{(2\pi)^d\mathbf{g}}\frac{e^{-iqx}}{\mathbf{r}\mathbf{g}^{-1} + q^2}
\end{align}

The right side of (\ref{2_point_sad}) is just the inverse Fourrier transform of the Lorenz function, and hence we obtain

\begin{equation}
C_{kl} \sim e^{-x\sqrt{\mathbf{g}/\mathbf{r}}}
\end{equation}

Our main goal though, is to reach a state of self-similarity, when the system displays scale-invariance; this is the whole scope of the RG procedure, resulting in the equations (\ref{couplings_rg}). 
In this case, $\mathbf{r}\rightarrow 0$ and the correlation function (\ref{2_point}) simplifies to

\begin{align}
\label{2_point_crit}
C_{kl} = \langle\phi_k\phi_l\rangle_0 = \beta \int \frac{d^dq}{(2\pi)^d}\frac{e^{-i\mathbf{q}\mathbf{x}}}{\mathbf{g}q^2}\sim \frac{1}{|\mathbf{x}|^{d-2}}
\end{align}

For our case of interest when $d=2$, the integral diverges as $\ln|\mathbf{x}|$, hence the long range correlation.



\end{document}